\documentclass[12pt]{iopart}

\jl{6}   % Submit to CQG

\newcommand{\lie}{{\cal L}_\beta}
\newcommand{\tder}{\left( \partial_t - \lie \right)}
\newcommand{\Gt}{\tilde \Gamma^i}

\begin{document}

\title[]{The BSSN formulation is a partially constrained evolution system}

\author{Adrian P Gentle}

\address{Department of Mathematics, University of Southern Indiana,
  Evansville, IN 47712}

\ead{apgentle@usi.edu}

\date{\today}

\begin{abstract} 
  Relativistic simulations in 3+1 dimensions typically monitor the
  Hamiltonian and momentum constraints during evolution, with
  significant violations of these constraints indicating the presence
  of instabilities.  In this paper we rewrite the momentum constraints
  as first-order evolution equations, and show that the popular BSSN
  formulation of the Einstein equations explicitly uses the momentum
  constraints as evolution equations.  We conjecture that this feature
  is a key reason for the relative success of the BSSN formulation in
  numerical relativity.
\end{abstract}

%\submitto{\CQG} 
\pacs{04.20.-q, 04.25.Dm}

\section{Introduction}

Three decades after the earliest attempts to study black hole
spacetimes numerically, the problem of simulating the collision and
coalescence of two black holes remains at the forefront of numerical
relativity.  Only in the last year have the first stable, long-term
numerical simulations of black hole collision and coalescence in
three-dimensions been achieved \cite{baker06,campanelli06}.  The
primary focus of numerical relativity has at last shifted to the
astrophysical implications of these events, with recent simulations
investigating recoil in the final product of binary black hole
mergers, and the extraction of gravitational wave profiles for
comparison with observations \cite{baker06b,gonzalez07,laguna07}.

Despite these advances, progress is still needed in understanding both
the mathematical and computational formulations of the equations.  The
most popular formulation of the Einstein equations in numerical
relativity is the BSSN system \cite{BSSN0,BSSN1} (but see
\cite{pretorius} for an alternate approach).  The Einstein equations
are split into distinct evolution and constraint equations, with the
constraints being used to construct initial data, after which they are
monitored (rather than enforced) as the evolution proceeds.  In
addition, a variety of technical and computational techniques are
employed to achieve stable evolutions, including causal differencing,
artificial diffusion, grid excision, moving punctures, special shift
conditions, and many others which have been tried over the last few
decades.

In this paper we consider the role played by the momentum constraints
in evolution, and in particular, their appearance in the BSSN
formulation.  It is well known that the momentum constraints are vital
to the stability of the BSSN system, where they are used to eliminate
the spatial divergence of the trace-free extrinsic curvature from one
of the evolution equations.  This is a necessary condition for
stability \cite{BSSN1}.  The Hamiltonian and momentum constraints are
then monitored during evolution, and give an indication of the quality
and stability of the simulation.

The close link between unstable simulations and uncontrolled violation
of the constraints has led to the development of several constrained
evolution algorithms
\cite{bonazzola04,anderson05,marronetti05,marronetti06}.  These
schemes are designed to enforce the Hamiltonian and/or momentum
constraints throughout the evolution, either through direct projection
or gentle numerical relaxation onto the constraint surface.

We will show that the BSSN system is already a partially constrained
formulation of the Einstein equations, since the evolution equations
used for the conformal connection functions are the momentum
constraints rewritten as first-order evolution equations.  It is in
this sense that the BSSN formulation is a partially constrained
evolution scheme, with a form of the momentum constraints being
actively enforced throughout the evolution.

We begin by briefly reviewing the BSSN formulation and its derivation,
highlighting the role played by the momentum constraints.  In section
3 the momentum constraints are recast as first-order evolution
equations for a particular combination of the metric and its spatial
derivatives.  Then, in section 4, a conformal transformation on this
variable demonstrates that the momentum constraints may be used to
evolve the conformal connection functions.  A more direct calculation
demonstrates that the resulting equation is identical to to the
standard BSSN evolution equation for $\Gt$.  In the final section we
consider the consequences of this observation.

\section{The ADM and BSSN evolution equations}

The Baumgarte-Shapiro-Shibata-Nakamura (BSSN) formulation of the
Einstein equations \cite{BSSN0,BSSN1} can be derived from the
canonical ADM form of the equations.  With the metric in the 3+1
form
\begin{equation}
  ds^2 = - ( \alpha^2 - \beta_i \beta^i )\, dt^2 + 2 \beta_i\, dt\, dx^i
  + g_{ij}\, dx^i dx^j,
\end{equation}
the components of the three-metric $g_{ij}$ become the dynamical
variables, and the remaining coordinate freedoms are expressed by the
lapse function $\alpha$ and shift vector $\beta^i$.  The Einstein
equations consist of six second order (in both time and space),
non-linear partial differential equations, together with an additional
four constraints which only contain first temporal derivatives of the
$g_{ij}$.  By introducing the extrinsic curvature tensor $K_{ij}$ the
Einstein equations can be written as a set of twelve evolution
equations which are explicitly first-order in time (neglecting source
terms for simplicity),
\begin{eqnarray}
  \label{eqn:adm1}
  \tder g_{ij} & = & - 2 \alpha K_{ij} \\
  \label{eqn:adm2}
  \tder K_{ij} & = & \alpha \left( R_{ij} - 2 K_{il} K^l_{\ j} + K K_{ij} \right)   - \nabla_i \nabla_j \alpha,
\end{eqnarray}
where $\lie$ is the Lie derivative along $\beta^i$ and
$K=g^{ij}K_{ij}$.  The constraint equations are
\begin{eqnarray}
  \label{eqn:hamiltonian}
  R + ( K)^2 - K_{ij} K^{ij}  = 2\rho,   \\ 
  \label{eqn:momentum}
  \nabla_j K^{ij} - g^{ij}\nabla_j  K = S^i,
\end{eqnarray}
and are known as the Hamiltonian and momentum constraints, respectively. These
are mathematically conserved by the evolution equations, in the sense
that if they are satisfied on some spacelike hypersurface then
solutions of the evolution equations will continue to satisfy the
constraints for all time.

This 3+1 form of the Einstein equations is known as the ADM system,
and was the basis of initial efforts to construct numerical solutions.
Until quite recently all attempts to evolve black holes in three
dimensions were unsuccessful (see \cite{lehner01} for a review), with
the impending failure of the code signposted by rapid growth in
violations of the constraints.  This led many investigators to propose
and test new formulations of the equations, among the most popular
being the so-called BSSN system developed by Shibata and Nakamura
\cite{BSSN0}, and later popularized by Baumgarte and Shapiro
\cite{BSSN1}.  Their approach splits the extrinsic curvature tensor
into its trace ($ K$) and trace-free ($A_{ij}$) parts, and then
performs a conformal decomposition of the three-geometry, with $g_{ij}
= e^{4\phi}\tilde g_{ij}$ and $A_{ij} = e^{ 4\phi}\, \tilde A_{ij}$.
Then
\begin{equation}
  \label{eqn:conformalK}
  K_{ij} = e^{4\phi}\, \left( \tilde A_{ij} + \frac{1}{3}\, \tilde g_{ij}\,  K \right),\\  
\end{equation}
and for consistency we require that $\det(\tilde g_{ij}) = 1$ and $ \tilde
A^i_{\ i} = \tilde g^{ij} \tilde A_{ij} = 0$.

The ADM evolution equations (\ref{eqn:adm1}) and (\ref{eqn:adm2}) are
transformed to evolve $\tilde g_{ij}$ and $\tilde A_{ij}$,
\begin{eqnarray}
\label{eqn:bssn1}  \tder \tilde g_{ij} & = & - 2 \alpha \tilde A_{ij} \\
\label{eqn:bssn2}  \tder \tilde A_{ij} & = & e^{-4\phi} \left[ 
    \alpha R_{ij} - \nabla_i \nabla_j \alpha \right]^{\mbox{\small TF}} +\ \alpha \left( 
  K\,\tilde A_{ij} - 2 \tilde A_{ik} \tilde A^{k}_{\ j} \right)
\end{eqnarray}
where ``TF'' denotes the trace-free portion of the bracketed
expression.  Similarly, the ADM equations lead to
\begin{eqnarray}
\label{eqn:bssn3}
  \tder \phi &=& -\frac{1}{6}\,\alpha\, K\\
\label{eqn:bssn4}
  \tder K &=& \alpha \left( \tilde A_{ij} \tilde A^{ij} 
    + \frac{1}{3}\, K^2 \right) - \nabla_i \nabla^i \alpha 
\end{eqnarray}
which are used to evolve {\large $\phi$} and {\large $ K$}.

The BSSN approach also introduces the conformal connection functions
\begin{equation}
\label{eqn:conf}
\tilde \Gamma^i = - \partial_j\, \tilde g^{ij},
\end{equation}
for which evolution equations can be obtained by differentiating with
respect to the time coordinate and
%\[
%\partial_t \tilde \Gamma^i = - \partial_t \partial_j \tilde g^{ij}
%\]
commuting the partial derivatives.  This yields
\begin{equation}
  \label{eqn:almostevn}
  \partial_t \tilde \Gamma^i = - \partial_j\, ( 2\alpha \tilde A^{ij} + \lie \tilde g^{ij} ) 
\end{equation}
where equation (\ref{eqn:bssn1}) has been used to eliminate explicit time
derivatives of the conformal three-metric.  The Lie derivative is given by
\begin{equation}
  \lie \tilde g^{ij} = \beta^k\partial_k \tilde g^{ij}   - 2 \tilde g^{k(i}\partial_k \beta^{j)} 
   + \frac{2}{3}\, \tilde g^{ij} \partial_k \beta^k,
\end{equation} 
since $\tilde g^{ij}$ is a tensor density of weight $\frac{2}{3}$.

In principle this completes the system of evolution equations, since
the conformal connection coefficients could be evolved using equation
(\ref{eqn:almostevn}).  However, it is found that the system is only
more stable than the original ADM formulation if the spatial
divergence of $\tilde A^{ij}$ is eliminated from equation
(\ref{eqn:almostevn}).  This is achieved using the momentum
constraints \cite{BSSN1}.  Under the conformal transformation
described above, the momentum constraints take the form
\begin{equation}
  \label{eqn:momconst}
\partial_j \tilde A^{ij}   
+ 6 \tilde A^{ij} \partial_j \phi
- \frac{2}{3}\, \tilde g^{ij} \partial_j K
+ \tilde \Gamma^i_{jk} \tilde A^{jk}  = e^{4\phi} S^i,
\end{equation}
where $S^i$ is a source term.  Using these to eliminate the spatial
divergence of $\tilde A^{ij}$ from (\ref{eqn:almostevn}) gives
\begin{eqnarray}
  \label{eqn:bssn5}
  \partial_t\, \tilde \Gamma^i & = & - 2 \tilde A^{ij} \partial_j \alpha
  + 2 \alpha \left( 6 \tilde A^{ij}\, \partial_j \phi  - \frac{2}{3}\, \tilde g^{ij}\, \partial_j  K 
    + \tilde \Gamma^i_{jk}\, \tilde A^{jk} - \tilde S^i\right)  \\
 & &  - \partial_j  \left( \beta^k\partial_k \tilde g^{ij}   - 2 \tilde g^{k(i}\partial_k \beta^{j)} 
   + \frac{2}{3}\, \tilde g^{ij} \partial_k \beta^k \right).  \nonumber
\end{eqnarray}

The standard BSSN evolution equations consist of
(\ref{eqn:bssn1})-(\ref{eqn:bssn4}) together with (\ref{eqn:bssn5}),
and are used to evolve the BSSN variables $\tilde g_{ij}$, $\tilde
A_{ij}, \phi, K$ and $\Gt$.  This system has been found to be
relatively stable, and is basis for most of the successful recent
binary black hole calculations \cite{baker06,campanelli06,laguna07}.

\section{The momentum constraints as first-order evolution equations }

The momentum constraints (\ref{eqn:momentum}) are a set of three
differential constraints relating the three-metric and extrinsic
curvature.  Since the extrinsic curvature is itself related to the
time development of the three-metric through its definition, equation
(\ref{eqn:adm1}), the momentum constraints can be viewed as equations
involving the time derivative of $g_{ij}$.  In this section we
explicitly rewrite the momentum constraints as first-order evolution
equations for a function of the three-metric and its spatial
derivatives.

The momentum constraints with source terms, equation (\ref{eqn:momentum}), can be written as
\begin{equation}
  \label{eqn:expandmom}
  \partial_j K^{ij} + \Gamma^j_{jk} K^{ik} + \Gamma^i_{jk} K^{jk} - g^{ij}\partial_j  K = S^i,
\end{equation}
where $\Gamma^i_{jk}$ are the Christoffel symbols and $S^i$ represents the
sources.  Using the definition of extrinsic curvature, equation
(\ref{eqn:adm1}), the first term in the constraints becomes
\[ 
\partial_j K^{ij} = \partial_j \left(
  \frac{1}{2 \alpha} \left( \partial_t g^{ij} - \lie g^{ij}  \right) \right),
\]
and by commuting the spatial and temporal partial derivatives of $g_{ij}$ we find that 
\[
\partial_j K^{ij} = \frac{1}{2\alpha} \biggl( \partial_t \left( \partial_j g^{ij} \right)
   - 2 K^{ij} \partial_j \alpha - \partial_j \left(\lie g^{ij}  \right) \biggr).
\]
Applying a similar procedure to the gradient of $K$, we have
\begin{eqnarray*}
g^{ij}\partial_j K & = &  \frac{1}{2\alpha} \biggl( 
- \partial_t \left( g^{ij}g^{ab}\partial_j g_{ab} \right)
- 2 K g^{ij} \partial_j \alpha
+ g^{ab} K^{ij} \partial_i g_{ab}\\
& & + g^{ij} g^{ab}\partial_j \left( \lie g_{ab} \right)
+ g^{ij} \partial_j g_{ab} \lie g^{ab}
+ g^{ab} \partial_j g_{ab} \lie g^{ij}
\biggr).
\end{eqnarray*}
Combining these results we can write the momentum constraints
(\ref{eqn:expandmom}) as
\begin{eqnarray*}
  \partial_t \left( \partial_j g^{ij} + g^{ij}  \partial_j \ln g \right) & = & 
   2 \left( K^{ij}  - g^{ij} K \right) \partial_j \alpha \\
  & & + 2 \alpha \left(  K^{ij} \partial_j \ln g + 
    S^{i} - \Gamma^j_{jk} K^{ik} - \Gamma^i_{jk} K^{jk} \right) \\
  & & + \partial_j \left( \lie g^{ij} \right)
  + g^{ij} \partial_j \left( g^{ab} \lie g_{ab} \right)
  + (\partial_j\ln g)\, \lie g^{ij}
\end{eqnarray*}
where we have used
$g^{ab} \partial_j g_{ab} =  \partial_j \ln g$ and $g=\det(g_{ij})$.

The momentum constraints can thus be viewed as the natural evolution
equations for the subsidiary variable
\[
\Lambda^i = \partial_j g^{ij} + g^{ij} \partial_j \ln g, 
\]
the first term of which is suggestive of the conformal connection
coefficients defined by equation (\ref{eqn:conf}).  The momentum
constraints written in their most general form as evolution equations
are then
\begin{eqnarray}
  \label{eqn:lambdaevolve}
  \partial_t \Lambda^i & = & 
   2 \left( K^{ij}  - g^{ij} K \right) \partial_j \alpha  + 2 \alpha \left(  K^{ij} \partial_j \ln g + 
    S^{i} - \Gamma^j_{jk} K^{ik} - \Gamma^i_{jk} K^{jk} \right) \nonumber \\
  & & + \partial_j \left( \lie g^{ij} \right)
  + g^{ij} \partial_j \left( g^{ab} \lie g_{ab} \right)
  + (\partial_j\ln g)\, \lie g^{ij},
\end{eqnarray}
where $\Lambda^i$ is a subsidiary variable since it is not truly
independent, but rather a function of the metric and its spatial
derivatives.  We conjecture that using the momentum constraints to
evolve $\Lambda^i$ in the ADM formulation, alongside the evolution of
$g_{ij}$ and $K_{ij}$, would enforce the consistency of the
constrained evolution system more thoroughly than evolving the
three-metric and extrinsic curvature alone.

Finally, we note that equation (\ref{eqn:lambdaevolve}) is essentially
the same form of the momentum constraint used in the first-order
flux-conservative formulation of Bona and Masso \cite{arbona99}.
The Bona-Masso approach introduces a new variable $V_i$ which is
evolved using the momentum constraint, and in our notation $V_i =
\frac{1}{2}\, g_{ij} \Lambda^j.$

\section{Momentum constraints in the BSSN formulation}

In the previous section we demonstrated that the momentum constraints
can be used to evolve the subsidiary variable $\Lambda^i$, which is
itself a function of the metric and its spatial derivatives.  In this
section we show that under a conformal transformation $\Lambda^i$ is
closely related to the BSSN variable $\tilde \Gamma^i$.  Guided by
this insight, we will derive the evolution equation for $\tilde
\Gamma^i$ directly from the conformally decomposed momentum
constraints.

Beginning with the variable $\Lambda^i$ introduced above, the conformal
decomposition used in section 2 gives
\begin{eqnarray*}
\Lambda^i  = 
 e^{-4 \phi}\left(  \partial_j\, \tilde g^{ij} 
 - \, 8\, \tilde g^{ij}\, \partial_j\, \phi \right),
 \end{eqnarray*}
 which suggests that we define $\Lambda^i = e^{-4\phi} \tilde
 \Lambda^i$, so that
 \begin{eqnarray*}
\tilde \Lambda^i  =    - \tilde \Gamma^i - 8 \, \tilde g^{ij}\, \partial_j \phi.
\end{eqnarray*}
It is clear that $\tilde \Lambda^i$ splits into the
BSSN conformal connection coefficient and the gradient of the conformal
factor.

It follows from the results of the previous section that the momentum
constraint can be used to evolve $\Gt$.  By calculating the
time derivative of $\Lambda^i$ we find that
\[
\partial_t \Lambda^i = 
e^{-4\phi} \biggl( 
\partial_t \tilde \Gamma^i - 4\, \partial_t \left(
  \tilde g^{ij} \partial_i \phi \right) +\ 2 \left( \partial_i \tilde
  g^{ij} + 8 \tilde g^{ij}\partial_i \phi\, \right) \partial_t\phi
\biggr),
\]
and eliminating the $\partial_t \phi$ and $\partial_t \tilde g^{ij}$
terms, using equations (\ref{eqn:bssn1}) and (\ref{eqn:bssn3}),
relates the time evolution of $\Lambda^i$ directly to the evolution of
$\Gt$.  Substituting this into the momentum constraints in the form of
equation (\ref{eqn:lambdaevolve}) will then give an evolution equation
for $\tilde \Gamma^i$.

This shows, somewhat indirectly, that the momentum constraints can be
used to evolve the conformal connection coefficients $\Gt$.  It is
more enlightening, however, if we begin with the conformal form of the
momentum constraints,
\begin{equation}
  \label{eqn:momconst2}
\partial_j \tilde A^{ij}   
+ 6 \tilde A^{ij} \partial_j \phi
- \frac{2}{3}\, \tilde g^{ij} \partial_j K 
+ \tilde \Gamma^i_{jk} \tilde A^{jk} = e^{4\phi} S^i
\end{equation}
and perform a calculation analogous to that of the previous section.

It follows directly from the definition of $\tilde A_{ij}$, equation
(\ref{eqn:bssn1}), that
\[
\tilde A^{ij} = \frac{1}{2\alpha}\left( \partial_t \tilde g^{ij} 
- \lie \tilde g^{ij} \right).
\]
This allows us to re-express the first term in the momentum constraints
as
\[
\partial_j \tilde A^{ij} = \frac{1}{2\alpha} \biggl( \
 \partial_j \partial_t \tilde g^{ij}
- \partial_j ( \lie \tilde g^{ij} )\  \biggr)
-  \frac{1}{\alpha}\, \tilde A^{ij} \partial_j \alpha,
\]
and commuting the partial derivatives of $\tilde g^{ij}$ and using the
definition of $\Gt$ we can write
\[
\partial_j \tilde A^{ij} = - \frac{1}{2\alpha} \left( \
 \partial_t \tilde \Gamma^i
+  2 \tilde A^{ij} \partial_j \alpha 
+ \partial_j ( \lie \tilde g^{ij} )\  \right).
\]
Using this result to replace the divergence of $\tilde A^{ij}$ in the
conformal momentum constraints, equation (\ref{eqn:momconst2}), we
find
\begin{eqnarray}
  &- \frac{1}{2\alpha} \left( \
    \partial_t \tilde \Gamma^i
    +  2 \tilde A^{ij} \partial_j \alpha 
+ \partial_j ( \lie \tilde g^{ij} )\  \right) \nonumber \\
& \qquad\qquad 
+ 6 \tilde A^{ij} \partial_j \phi
- \frac{2}{3}\, \tilde g^{ij} \partial_j K 
+ \tilde \Gamma^i_{jk} \tilde A^{jk} 
= e^{4\phi} S^i. \nonumber
\end{eqnarray}
It is important to note that these are precisely the momentum
constraints; all that we have done is replace $\tilde A^{ij}$ with its
definition in terms of $\tilde g^{ij}$. Finally, expanding the Lie
derivative and rearranging yields the momentum constraints in a new
form:
\begin{eqnarray*}
    \partial_t\, \tilde \Gamma^i & = & - 2 \tilde A^{ij} \partial_j \alpha
  + 2 \alpha \left( 6 \tilde A^{ij}\, \partial_j \phi  - \frac{2}{3}\, \tilde g^{ij}\, \partial_i  K 
    + \tilde \Gamma^i_{jk}\, \tilde A^{jk} - \tilde S^i \right)  \\
 & &  - \partial_j  \left( \beta^k\partial_k \tilde g^{ij}   - 2 \tilde g^{k(i}\partial_k \beta^{j)} 
   + \frac{2}{3}\, \tilde g^{ij} \partial_k \beta^k \right).  \nonumber
\end{eqnarray*}

This is identical to equation (\ref{eqn:bssn5}), the standard equation
used to evolve the conformal connection coefficients in the BSSN
formulation.  From this we conclude that the usual statement that the
BSSN formulation is stable when the spatial divergence of $\tilde
A^{ij}$ is eliminated from the evolution equation for $\Gt$
\cite{BSSN1} is equivalent to the statement that {\em the BSSN system
is stable when the momentum constraints themselves are used to evolve
the conformal connection coefficients}.

\section{Conclusions}

We have shown that the BSSN formulation utilizes the momentum
constraints as evolution equations for the conformal connection
functions $\tilde \Gamma^i$.  It follows that the momentum constraints
should be satisfied to computational accuracy throughout the
evolution, although in practice violations of the ``definitional
constraints''
\begin{equation}
\label{eqn:newconst}
\tilde\Gamma^i + \partial_j\, \tilde g^{ij} = 0, \qquad 
\det(\tilde g_{ij}) =1, \qquad \mbox{and} \qquad
\tilde g^{ij} \tilde A_{ij} = 0,
\end{equation}
could lead to errors when the momentum constraints
 (\ref{eqn:momconst2}) are calculated during evolution.

We conjecture that the BSSN formulation shows superior stability
properties to the ADM system because errors in the definitional
constraints do not correspond directly to spurious sources of momentum
being pumped into the simulation.  Although the Hamiltonian and
momentum constraints are mathematically conserved by the ADM evolution
equations, small computational errors prevent the strict consistency
of the system.  The particular advantage of the BSSN system is that
the conformal connection functions provide a way of more strictly and
actively enforcing the consistency of the equations, provided the
violation of the Hamiltonian constraint is not too great.

We expect that in practice any violation observed when calculating the
momentum constraints (\ref{eqn:momconst2}) during evolution arises from
inconsistencies introduce by violations of the new definitional
constraints (\ref{eqn:newconst}).  It is possible that additional
stability in the system may be gained by including the addition terms
in the evolution equation for $\tilde \Gamma^i$ which ``drop out'',
but may in practise be non-zero.  For example, terms containing
\[
\tilde g^{ab} \partial_j \tilde g_{ab} \equiv \partial_j \ln \tilde g
\]
are nominally zero because $\tilde g=\det(\tilde g_{ij}) =1$, but in
practise we expect small violations in one of more of the constraints
(\ref{eqn:newconst}).

Our result is consistent with recent work by Marronetti
\cite{marronetti05,marronetti06}, where a constraint relaxation method
is applied to the BSSN formulation.  Although weak enforcement of the
Hamiltonian constraint was found to provide significant improvement in
the results \cite{marronetti05}, there was little change when a
similar approach was applied to the momentum constraints
\cite{marronetti06}.  As we have seen, the momentum constraints are
already being directly enforced in the BSSN approach.  It is likely
that the small improvements seen in Marronetti's momentum relaxation
scheme were due to improved (albeit indirect)
enforcement of the remaining constraints $\Gt + \partial_j \tilde
g^{ij}=0$ and $\det(\tilde g_{ij}) = 1$.

We have shown that the BSSN formulation ensures that the momentum
constraints are enforced throughout the evolution.  The generalized
harmonic approach of Pretorius \cite{pretorius} is also, by
construction, constraint damping.  Since all of the successful codes
currently used in numerical relativity are based on one of these
formulations, it appears that active constraint enforcement is a vital
ingredient for the success of any new formulation of the Einstein
equations designed for use in numerical relativity.

\section*{Acknowledgments} 
The author is grateful to the Lilly Endowment and the Pott Foundation
for support through their Summer Research Fellowship programs, and to
the School of Mathematical Sciences at Monash University for its
hospitality while some of this work was undertaken.  The author is
indebted to many colleagues for insightful comments and conversations,
including Pablo Laguna, Leo Brewin, Tony Lun, and Robert Bartnik.

% --- REFERENCES ----------------------------------------------------

\end{document}